\documentclass[twocolumn,prb]{revtex4}
\usepackage[T1]{fontenc}
\usepackage[latin1]{inputenc}
\usepackage{graphicx}
\usepackage{amssymb}


\begin{document}

\title{Resonant magnetic X-ray scattering spectra in SDW Cr\\
-- ab initio study --}

\author{Manabu Takahashi}

\affiliation{Faculty of Engineering, Gunma Univ, Kiryu, Gunma 376-8515, Japan}

\author{Jun-Ichi Igarashi}

\affiliation{Faculty of Science, Ibaraki Univ., Mito, Ibaraki 310-8512, Japan}

\author{Kunitomo Hirai}

\affiliation{Department of Physics, Nara Med. Univ., Kashiwara, Nara 634-8521,
Japan}

\begin{abstract}
Using \emph{ab-initio} band structure calculation based on the local
density approximation, Cr $K$-edge resonant X-ray magnetic scattering
spectra are analyzed in the spin density wave (SDW) state of chromium.
We perform band structure calculation, assuming an ideal bcc lattice
structure with the lattice constant observed at the spin-flip temperature
$T_{\mathrm{SF}}$ and a commensurate SDW state with the propagation
vector close to the observed value at $T_{\mathrm{SF}}$. Taking account
of the spin-orbit interaction, we obtain the orbital moment on each
Cr site induced in proportion to the local spin moment, which is quite
small, at most a tenth of those in nickel or iron. In spite of the
tiny $3d$ orbital moment, the orbital polarization is found to have
large fluctuations as a function of energy. We obtain the scattering
intensity at the Cr $K$-edge on the SDW magnetic Bragg spot, which
shows resonant enhancement in good agreement with the experiment.
The $3d$ orbital polarization is found to be highly correlated with
the intensity of the resonant main peak, indicating that the $4p$
orbital polarization is mainly induced by the $3d$ orbital polarization
through the $p$-$d$ hybridization.
\end{abstract}

\pacs{78.70.Ck, 75.30.Fv, 75.25.+z, 71.20.Be, 71.20.-b}

\maketitle

\section{Introduction}

Resonant X-ray diffraction technique has attracted much attention
in the field of studying spin and orbital properties of the $3d$
or $4f$ electrons in transition metal or rear-earth metal compounds.
Both resonant and non-resonant diffraction techniques have been revealing
the charge, spin, orbital and lattice characters in a wide variety
of compounds. In particular magnetic diffraction techniques have been
applied to various magnetic materials for studying the spin and orbital
magnetic properties.\cite{deBergevin1981,Namikawa1985,Blume1988,Hill1997,Gibbs1988,Hannon1988,Carra1989,Neubeck1999,Stunault1999,Neubeck2001,Mannix2001,Mannix2001b,Caciuffo2002,Paolasini2002,Wilkins2003,Veenendaal2003}
In this paper we focus on the magnetic diffraction in the spin density
wave in chromium.\cite{Hill1997,Mannix2001}

Chromium is a prototypical itinerant antiferromagnet forming a spin
density wave (SDW) state below the N\'eel temperature. A lot of studies
have been carried out for revealing the electronic and magnetic properties
of chromium.\cite{Fawcett1988,Mori1993,Hill1997,Mannix2001,Evans2002}
Hill et al.\cite{Hill1997} studied the SDW state and the charge density
wave (CDW) states by means of the X-ray scattering at the Cr $K$-edge.
They estimated the lattice distortion by analyzing the nonresonant
intensity, but failed to observe resonant enhancement at the Cr $K$-edge.
On the other hand Mannix et al.\cite{Mannix2001} have successfully
observed clear resonant enhancement at the Cr $K$-edge and analyzed
the spectra using an atomic model. Noting that the resonant enhancement
of the magnetic scattering intensity is mainly caused by the orbital
polarization (OP) in the $4p$ band,\cite{Hannon1988} %
\footnote{Orbital polarization here is concerned with the orbital angular momentum.%
}it is interesting that a clear resonant peak is observed in spite
of the very small orbital moment in the ground state.\cite{Mannix2001} 

We analyze the resonant X-ray magnetic scattering spectra in the SDW
chromium, using an \emph{ab initio} band structure calculation based
on the local density approximation (LDA). We cannot rely on the tight-binding
model for the $4p$ band, because the $4p$ wave function is greatly
extended in space so that in the crystal most weight of the $4p$
wave function is distributed on the neighboring site or interstitial
region. We first carry out a conventional band calculation without
taking account of the spin-orbit interaction (SOI). We obtain the
ground state potential which gives the SDW state, then calculate the
wave function in conduction band under the potential. This calculation
reproduces the same results as those already reported.\cite{Hirai1998,Hafner2002}
We take account of the SOI in the final stage. We find that the orbital
moment is induced in proportion to the local spin moment on each site,
although the orbital moment is as small as a tenth of those in ferromagnetic
Fe, Co or Ni.\cite{Iga1994a,Iga1996b} It should be noted here that
the $3d$ OP fluctuates considerably as a function of energy in spite
of the tiny orbital moment in the ground state.

Using the wave function thus obtained within the LDA, we calculate
the scattering intensity. The calculation reproduces well the photon
energy dependence on the SDW Bragg spot and the Fano-type dip in the
lower energy side of the resonant main peak.\cite{Mannix2001} The
Fano effect originates from the interference between the amplitudes
of the resonant and nonresonant scattering processes. The intensity
of the resonant main peak arises from the OP in the $4p$ band. In
order to clarify how the $4p$ OP is induced, we calculate the spectra
with turning on and off the SOI selectively in the $3d$ and $4p$
states. We find that the $4p$ OP is induced by the $3d$ OP through
the $p$-$d$ hybridization, and that the $3d$ OP strongly correlates
with the resonant peak intensity.

This reminds us of the mechanism of the magnetic circular dichroism
(MCD) spectra of X-ray absorption at the transition metal $K$-edge.
We studied the MCD spectra in Fe, Co, Ni\cite{Iga1994a,Iga1996b}
and Mn$_{3}$GaC\cite{Taka2003b}, and revealed that the intensity
of the dichroism spectra correlates with the $3d$ OP through the
hybridization between the $4p$ and $3d$ states.\cite{Iga1994a,Iga1996b,Rueff1998,Taka2003b}
Also, the sensitivity to electronic states at neighboring sites is
closely related to the mechanism of the RXS intensity on the orbital
ordering Bragg spot in transition metal compounds. Detailed studies\cite{Elfimov1999,Benfatto1999,Taka1999,Taka2001,Taka2002,Taka2003a}
based on band structure calculations have revealed that the intensity
is mainly controlled by the lattice distortion through the hybridization
of the transition-metal $4p$ states and the oxygen $2p$ states at
neighboring sites. 

This paper is organized as follows. In the next section, we briefly
describe the band structure calculation and the ground state. In Sec.
III we formulate the magnetic scattering amplitude. We discuss the
photon energy dependence of the scattering intensity and its relation
to the electronic states in Sec. IV. The last section is devoted to
the concluding remarks.

\section{Band structure calculation}

Chromium forms a bcc lattice structure. As cooling its magnetic state
turns into a transverse spin density wave (TSDW) state at its N\'eel
temperature $T_{{\rm N}}=311\,{\rm K}$ and into a longitudinal spin
density wave (LSDW) state at its spin-flip temperature $T_{{\rm SF}}=122\,{\rm K}$.
In the TSDW (LSDW) state, magnetic moments are perpendicular (parallel)
to the SDW propagation vector. The wavelength of the SDW is incommensurate
with the lattice periodicity.\cite{Fawcett1988} Charge density wave
(CDW) and lattice strain wave (LSW) are accompanied by the SDW.\cite{Pynn1976,Fawcett1988,Hill1995,Mori1993}. 

\begin{figure}[h]
\begin{center}\includegraphics[%
  scale=0.4]{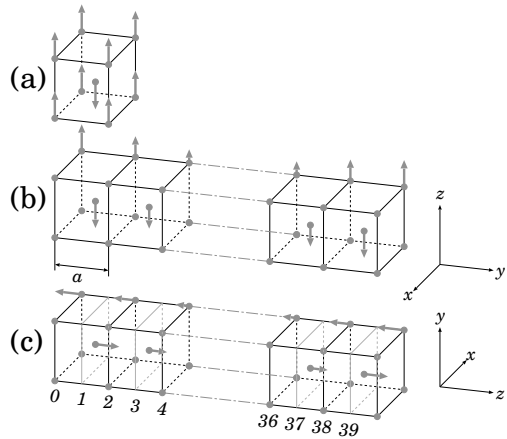}

\caption{Schematics of SDW states with $Q_{\mathrm{SDW}}=\frac{2\pi}{a}\frac{19}{20}$.
The arrows represent the direction and the magnitude of magnetic moment.
(a) a simple AF states. (b) a TSDW state. $a$ is the lattice parameter
of simple bcc structure. (c) an LSDW state. \label{cap:Schematic_View}}

\end{center}
\end{figure}

In the present calculation, we assume an ideal bcc lattice structure
with the lattice constant $a=5.45a_{0}$, where $a_{0}=0.529$\AA{}
is Bohr radius. The SDW wavelength is assumed to be $\lambda_{{\rm SDW}}=20a$,
which is very close to the observed value at the spin-flip temperature
$T_{{\rm SF}}$. Figure \ref{cap:Schematic_View} schematically depicts
the assumed magnetic structure. For convenience, the coordinate system
is chosen such that the $z$ direction is parallel to the magnetic
moment. We first carry out the band calculation selfconsistently without
the SOI, using the Korringa-Kohn-Rostoker (KKR) method with muffin-tin
(MT) approximation. The size of the MT sphere is chosen such that
neighboring spheres are touching each other. The maximum number of
$k$ points $n_{k}=n_{kx}\times n_{ky}\times n_{kz}$ is $30\times30\times2$
in the first Brillouin zone (FBZ) for the LSDW state and $16\times2\times16$
for the TSDW state. Then we calculate the eigenvalues and eigenfunctions
with adding the SOI term $\frac{1}{r}\frac{d}{dr}V\left(r\right)\ell_{z}s_{z}$
to the muffin-tin potential $V(r)$. Selfconsistent iteration is not
carried out. The one electron excited states are calculated up to
the energy $\epsilon_{\mathrm{F}}+2.0\,\mathrm{ryd}$, where $\epsilon_{\mathrm{F}}$
is the Fermi level. We neglect the spin flip terms $\ell_{+}s_{-}$
and $\ell_{-}s_{+}$ in the SOI. We expect that the neglected terms
cause only minor correction to the OP in transition metals and their
compounds.\cite{Taka2003a,Taka2003b,Usuda2004b} Since we are dealing
with a large unit cell containing 40 inequivalent Cr atoms, this may
help us to reduce the calculation volume.

We define the spin moment at the $j$th site as the difference between
electron numbers with spin up and spin down inside the MT sphere placed
at the $j$th site. That carried by the valence $d$ electrons $\mu_{\mathrm{spin}j}^{d}$
is defined as\begin{equation}
\mu_{\mathrm{spin}j}^{d}=\frac{1}{2}\left(n_{j\uparrow}^{d}-n_{j\downarrow}^{d}\right),\label{eq:spin_moment_def}\end{equation}
where $n_{j\alpha}^{d}$ is the number of spin $\alpha$ electrons
in the $d$ state inside the $j$th MT sphere, and those carried by
the valence $s$ and $p$ electrons are similarly defined. Another
important quantity is the local orbital moment at each site. That
carried by the valence $d$ electrons at the $j$th site $\mu_{\mathrm{orb}j}^{d}$
is defined as\begin{equation}
\mu_{\mathrm{orb}j}^{d}=\sum_{\alpha}\sum_{m=-2}^{2}n_{j\alpha}^{dm}m,\label{eq:orbital_moment_def}\end{equation}
where $m$ represents the magnetic quantum number, $n_{j\alpha}^{dm}$
is the number of the spin $\alpha$ electrons in the $d$ state specified
by $m$ inside the $j$th MT sphere. The quantity carried by the valence
$p$ electrons is similarly defined.

\begin{figure}[h]
\begin{center}\includegraphics[%
  scale=0.5]{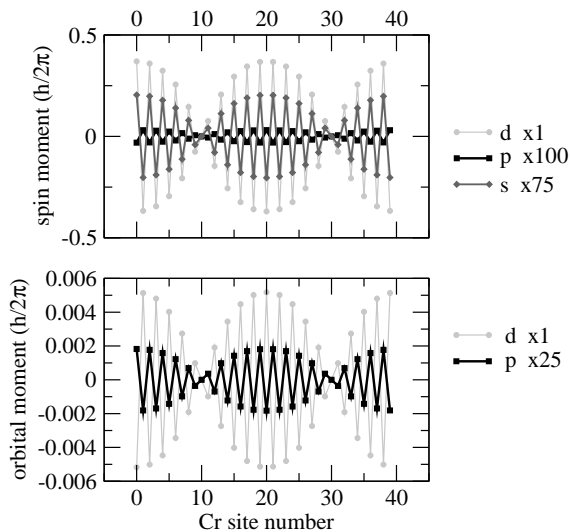}

\caption{Spin(upper panel) and orbital(lower panel) moment distribution in
LSDW state. The number of $k$ points in the FBZ is $16\times16\times2$.\label{cap:SOLSDW} }

\end{center}
\end{figure}

Figure \ref{cap:SOLSDW} shows the distribution of the local spin
and orbital moments in the LSDW state. The contributions of the valence
$s$ and $p$ electrons to the local spin moments are negligible compared
with those of the $d$ electrons. The local orbital moments are also
sinusoidally modulated, being proportional to the local spin moment.
The magnitude of the local orbital moment is quite small. The maximum
value is about $0.004\hbar$, being smaller than a tenth of those
in Fe, Co, and Ni.\cite{Iga1994a,Iga1996b} Therefore the contribution
of the orbital moment to the local moment is negligible. Note that
the spin and orbital moment distribution in the TSDW state are nearly
the same as those in the LSDW state, except for a slight reduction
of the $p$ orbital moment.

The moments thus obtained may be written as\begin{eqnarray}
\mu_{j} & = & M_{1}\cos\left(\mathbf{Q}_{\mathrm{SDW}}\cdot\mbox{\boldmath$\tau$}_{j}\right)+M_{3}\cos\left(3\mathbf{Q}_{\mathrm{SDW}}\cdot\mbox{\boldmath$\tau$}_{j}\right)\nonumber \\
 &  & +M_{5}\cos\left(5\mathbf{Q}_{\mathrm{SDW}}\cdot\mbox{\boldmath$\tau$}_{j}\right)+\cdots\label{eq:moment_fourie}\end{eqnarray}
in terms of the amplitude of the fundamental wave $M_{1}$ and the
amplitude of the odd-order harmonics $M_{3}$, $M_{5}$,$\cdots$,
where \textbf{$\mathbf{Q}_{\mathrm{SDW}}$} is a propagation vector
of the SDW, $\mathbf{Q}_{\mathrm{SDW}}=\mathbf{Q}_{\mathrm{TSDW}}=\frac{2\pi}{a}\left(0,\frac{19}{20},0\right)$
for the TSDW state and $\mathbf{Q}_{\mathrm{SDW}}=\mathbf{Q}_{\mathrm{LSDW}}=\frac{2\pi}{a}\left(0,0,\frac{19}{20}\right)$
for the LSDW state. $\mbox{\boldmath$\tau$}_{j}$ denotes the position
vector of the $j$th site, $\mbox{\boldmath$\tau$}_{j}=ja\hat{\mathbf{z}}\left(\mathbf{\hat{\mathbf{y}}}\right)$
for $n=0,2,4,\cdots$ and $\mbox{\boldmath$\tau$}_{j}=ja\hat{\mathbf{z}}\left(\mathbf{\hat{\mathbf{y}}}\right)+\frac{1}{2}a\left(\hat{\mathbf{x}}+\hat{\mathbf{y}}+\hat{\mathbf{z}}\right)$
in the LSDW (TSDW) state. For the $3d$ spin moments, the coefficients
in Eq. (\ref{eq:moment_fourie}) are evaluated as $M_{1}^{3d}=0.393\hbar$,
$M_{3}^{3d}=-0.026\hbar$, $M_{5}^{3d}=0.0025\hbar$ in the LSDW state.
These values are consistent with the reported band calculations\cite{Hirai1998,Hafner2002}
and the experiment.\cite{Mannix2001}

The charge density is also modulated, but its magnitude is underestimated
compared with the experimental value.\cite{Hill1997} If the LSW is
taken into account, this drawback may be improved.

\begin{figure}[h]
\begin{center}\includegraphics[%
  scale=0.6]{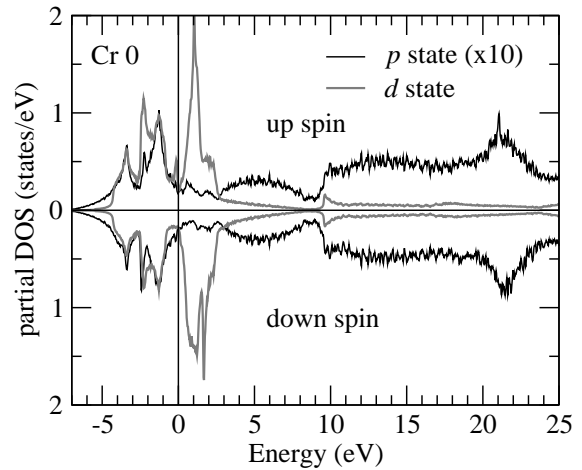}

\caption{Partial density of states projected on the MT sphere at $\mbox{\boldmath$\tau$}_{0}$.
Gray line and black line represent the DOS projected on the $d$ and
$p$ states, respectively. The number of $k$ points in the FBZ is
$30\times30\times2$. Fermi level $\epsilon_{\mathrm{F}}$ is at $0\,\mathrm{eV}$\label{cap:PDOS010}}

\end{center}
\end{figure}

Finally in this section, we discuss the partial density of states
(PDOS) defined as the density of states projected on the $p(\ell=1)$
or $d(\ell=2)$ state inside the MT sphere placed at each site. Figure
\ref{cap:PDOS010} shows those quantities at site $\mbox{\boldmath$\tau$}_{0}$,
which has the largest magnitude of spin moment. The PDOS in the TSDW
state is nearly the same as that in the LSDW state. The PDOS in the
previous band calculation\cite{Hafner2002} is limited in the energy
range $-7\sim5\,\mathrm{eV}$. The present result is essentially the
same as them. The PDOS of the $p$ state is significantly small compared
with that of the $d$ state. This is consistent with the fact that
the wave function of the $p$ state is greatly extended in space.

\section{Magnetic X-ray scattering spectra}

\subsection{Formulation}

\subsubsection{nonresonant magnetic scattering}

We use the formula given by Blume and Gibbs\cite{Blume1985,Blume1988}
for evaluating the nonresonant magnetic scattering amplitude. The
smallness of the orbital angular moment in Cr may allow to ignore
the orbital contribution to the scattering amplitude. Then we obtain
the expression of the amplitude $f_{\mathrm{NR}}$ as\begin{eqnarray}
 &  & f_{\mathrm{NR}}\left(\mathbf{q}^{\prime},\mathbf{e}^{\prime}\leftarrow\mathbf{q},\mathbf{e}\right)\nonumber \\
 &  & \,=NG\left(\mathbf{Q}\right)\sum_{j}e^{i\mathbf{Q}\cdot\mbox{\boldmath$\tau$}_{j}}f_{\mathrm{NR}}^{\mbox{\boldmath$\tau$}_{j}}\left(\mathbf{q}^{\prime},\mathbf{e}^{\prime}\leftarrow\mathbf{q},\mathbf{e}\right),\label{eq:nonreso_0}\end{eqnarray}
where\begin{eqnarray}
 &  & f_{\mathrm{NR}}^{\mbox{\boldmath$\tau$}_{j}}\left(\mathbf{q}^{\prime},\mathbf{e}^{\prime}\leftarrow\mathbf{q},\mathbf{e}\right)=A\left(-2i\frac{E_{{\rm Ryd}}}{\hbar\omega}\right)\nonumber \\
 &  & \,\times S_{z}^{\mbox{\boldmath$\tau$}_{j}}\left(\mathbf{Q}\right)\Bigl(\mathbf{e}^{\prime}\times\mathbf{e}-\left(\mathbf{e}^{\prime}\cdot\hat{\mathbf{q}}\right)\left(\hat{\mathbf{q}}\times\mathbf{e}\right)\nonumber \\
 &  & \,+\left(\mathbf{e}\cdot\hat{\mathbf{q}}^{\prime}\right)\left(\hat{\mathbf{q}}^{\prime}\times\mathbf{e}^{\prime}\right)-\left(\hat{\mathbf{q}}^{\prime}\times\mathbf{e}^{\prime}\right)\times\left(\hat{\mathbf{q}}^{\vphantom\prime}\times\mathbf{e}\right)\Bigr)_{z}.\label{eq:nonreso}\end{eqnarray}
Here, $f_{\mathrm{NR}}^{\mbox{\boldmath$\tau$}_{j}}$ is the scattering
amplitude of the atom at $\mbox{\boldmath$\tau$}_{j}$, and $N$ is
the number of unit cells in the crystal. Vectors $\mathbf{q}$ and
$\mathbf{e}$ represent the wave number and polarization of the incident
photon, respectively, while $\mathbf{q}^{\prime}$ and $\mathbf{e}^{\prime}$
are those of the emitted photon. Unit vectors $\hat{\mathbf{q}}$
and $\hat{\mathbf{q}}^{\prime}$ are defined as $\mathbf{q}/q$ and
$\mathbf{q}^{\prime}/q^{\prime}$, respectively. $\mathbf{Q}$ is
a scattering vector defined by $\mathbf{q}^{\prime}-\mathbf{q}$.
$\hbar\omega$, $E_{{\rm Ryd}}$, and $mc^{2}$ represent the photon
energy, the energy constant equal to $1\,\mathrm{Ryd}$, and the rest
energy of an electron, respectively. Function $G\left(\mathbf{Q}\right)$
is defined such that it is unity when $\mathbf{Q}$ coincides with
one of the reciprocal vectors for the SDW supper cell, and zero other
than the case. $S_{z}^{\mbox{\boldmath$\tau$}_{j}}\left(\mathbf{Q}\right)$
is a Fourier transform of the local spin moment inside the MT sphere
at $\mbox{\boldmath$\tau$}_{j}$. The direction of magnetization is
assumed to be parallel to the $z$ direction. The factor $A$ is given
by

\begin{equation}
A=\frac{\alpha^{2}}{4}\left(\frac{\hbar\omega}{E_{{\rm Ryd}}}\right)^{2},\label{eq:prefac_nonreso}\end{equation}
where $\alpha=\frac{e^{2}}{\hbar c}$ represents the fine structure
constant. 

\begin{figure}[h]
\begin{center}\includegraphics[%
  scale=0.5]{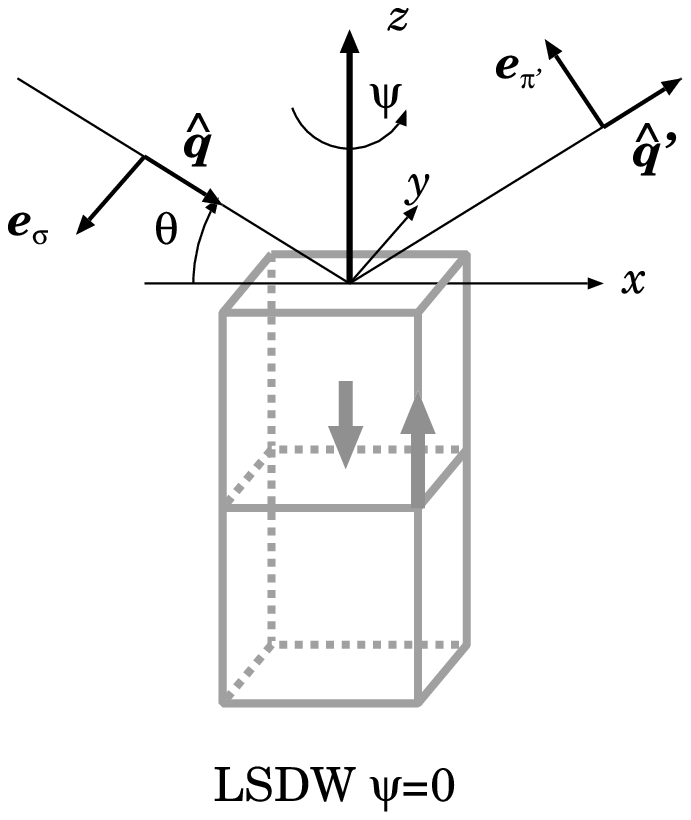} \includegraphics[%
  scale=0.5]{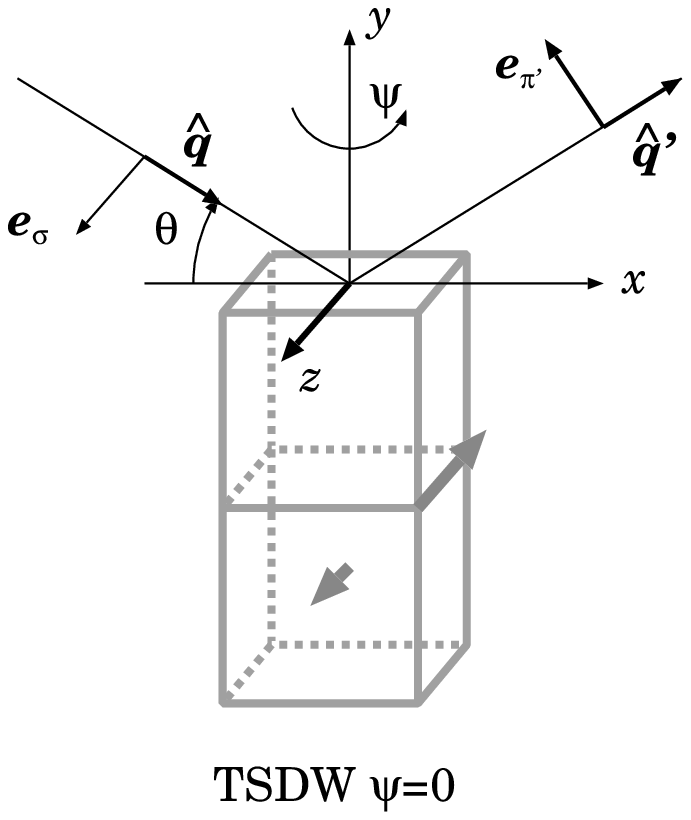}

\caption{Scattering geometry for the LSDW(left) and TSDW(right) configuration.
Gray arrows represent the direction of the magnetic moment.\label{cap:Geometry}}

\end{center}
\end{figure}

In the following, we consider the scattering in the $\sigma\pi^{\prime}$
channel with the scattering vector $\mathbf{Q}=\mathbf{Q}_{\mathrm{LSDW(TSDW)}}$
for the LSDW(TSDW) states. The scattering geometry is defined in Fig.
\ref{cap:Geometry}. The scattering angle $\theta$ and azimuthal
angle $\psi$ are shown. The scattering angle is $\theta\approx21\,\mathrm{deg.}$
in both of the LSDW and TSDW states. We can easily see that the scattering
amplitude for the $\sigma\sigma^{\prime}$ channel vanishes in the
LSDW state. The amplitude for the $\sigma\pi^{\prime}$ channel in
the LSDW state is given by \begin{eqnarray}
 &  & f_{\mathrm{NR}}\left(\mathbf{q}^{\prime},\mathbf{e}^{\prime}\leftarrow\mathbf{q},\mathbf{e}\right)=A\left(-\frac{2iE_{\mathrm{Ryd}}}{\hbar\omega}\right)2\sin^{3}\theta\tilde{S}_{z}\left(\mathbf{Q}\right)\nonumber \\
 &  & \qquad\qquad\qquad\approx-A\frac{i}{110}\sin^{3}\theta\tilde{S}_{z}\left(\mathbf{Q}\right),\label{eq:nonreso_lsdw}\end{eqnarray}
where \begin{equation}
\tilde{S}_{z}\left(\mathbf{Q}\right)=\sum_{j}e^{i\mathbf{Q}\cdot\mbox{\boldmath$\tau$}_{j}}S_{z}^{\mbox{\boldmath$\tau$}_{j}}\left(\mathbf{Q}\right).\label{eq:sum_spin_fourie}\end{equation}
Here we ignore the photon energy dependence assuming $\hbar\omega\approx5990\,\mathrm{eV}\approx440E_{\mathrm{Ryd}}$.
Also the amplitude for $\sigma\pi^{\prime}$ channel in the TSDW state
is given by \begin{eqnarray}
f_{\mathrm{NR}} & \approx & -A\frac{i}{110}\sin^{2}\theta\cos\theta\sin\psi\tilde{S}_{z}\left(\mathbf{Q}\right).\label{eq:nonreso_tsdw}\end{eqnarray}

Within the spherical approximation\cite{Coppens}, $S_{z}^{\mbox{\boldmath$\tau$}}\left(\mathbf{Q}\right)$
is expressed as 

\begin{eqnarray}
 &  & S_{z}^{\mbox{\boldmath$\tau$}_{j}}\left(\mathbf{Q}\right)=\frac{1}{2}\int_{0}^{R_{\mathrm{MT}}}\frac{\sin\left(Qr\right)}{Qr}r^{2}\nonumber \\
 &  & \qquad\qquad\qquad\times\left(\rho_{\mbox{\boldmath$\tau$}_{j}\uparrow}\left(r\right)-\rho_{\mbox{\boldmath$\tau$}_{j}\downarrow}\left(r\right)\right)dr,\label{eq:spherical_approx}\end{eqnarray}
where $R_{\mathrm{MT}}$ is the radius of the MT sphere and $\rho_{\mbox{\boldmath$\tau$}_{j}\alpha}\left(r\right)$
is the distribution function of the electrons with spin $\alpha$
inside the MT sphere at $\mbox{\boldmath$\tau$}_{j}$. With the help
of the band calculation, we evaluate Eq. (\ref{eq:spherical_approx})
and thereby Eqs. (\ref{eq:nonreso_lsdw}) and (\ref{eq:nonreso_tsdw}).

\subsubsection{resonant magnetic scattering}

We evaluate the resonant scattering amplitude exploiting the second
order Fermi's golden rule.\cite{Blume1985,Blume1988} We have three
terms, the orbital-orbital scattering term $f_{\mathrm{OO}}$, the
orbital-spin scattering term $f_{\mathrm{OS}}$, and the spin-spin
scattering term $f_{\mathrm{SS}}$.\cite{Blume1985,Blume1988} Since
$f_{\mathrm{OS}}\sim\frac{\hbar\omega}{mc^{2}}f_{\mathrm{OO}}\approx f_{\mathrm{OO}}\times10^{-2}$
and $f_{\mathrm{SS}}\sim\left(\frac{\hbar\omega}{mc^{2}}\right)^{2}f_{\mathrm{OO}}\approx f_{\mathrm{OO}}\times10^{-4}$
for the Cr $K$-edge, $f_{\mathrm{OS}}$ and $f_{\mathrm{SS}}$ can
be ignored in the present calculation. We consider only the dipole
process, since the quadrupole process gives rise to only minor contribution.

Accordingly the resonant scattering amplitude may be expressed as

\begin{eqnarray}
 &  & f_{\mathrm{R}}\left(\mathbf{q}^{\prime},\mathbf{e}^{\prime}\leftarrow\mathbf{q},\mathbf{e}\right)\nonumber \\
 &  & =NG\left(\mathbf{Q}\right)\sum_{j}e^{i\mathbf{Q}\cdot\mbox{\boldmath$\tau$}_{j}}f_{\mathrm{OO}}^{\mbox{\boldmath$\tau$}_{j}}\left(\mathbf{q}^{\prime},\mathbf{e}^{\prime}\leftarrow\mathbf{q},\mathbf{e}\right),\label{eq:amp_reso}\end{eqnarray}
where\begin{eqnarray}
 &  & f_{\mathrm{OO}}^{\mbox{\boldmath$\tau$}_{j}}\left(\mathbf{q}^{\prime},\mathbf{e}^{\prime}\leftarrow\mathbf{q},\mathbf{e}\right)\nonumber \\
 &  & =\frac{4\pi}{9}\sum_{m}\sum_{m^{\prime}}Y_{1m^{\prime}}\left(\mathbf{e}^{\prime}\right)Y_{1m}^{*}\left(\mathbf{e}\right)\nonumber \\
 &  & \times\sum_{n\mathbf{k}\alpha}A^{\prime}T_{n\mathbf{k}\alpha;\mbox{\boldmath$\tau$}_{j}1m^{\prime}\alpha}^{\dagger}T_{n\mathbf{k}\alpha;\mbox{\boldmath$\tau$}_{j}1m\alpha}\nonumber \\
 &  & \times\frac{F_{n\mathbf{k}\alpha}^{\mbox{\boldmath$\tau$}_{j}1m^{\prime}}F_{n\mathbf{k}\alpha}^{\mbox{\boldmath$\tau$}_{j}1m*}}{\left(\hbar\omega-\epsilon_{n\mathbf{k}\alpha}-\epsilon_{1s\alpha}+i\Gamma\right)/mc^{2}}\theta\left(\epsilon_{n\mathbf{k}\alpha}-\epsilon_{\mathrm{F}}\right),\label{eq:formfac_reso}\end{eqnarray}
and\begin{equation}
F_{n\mathbf{k}\alpha}^{\mbox{\boldmath$\tau$}_{j}\ell m}=\int\psi_{n\mathbf{k}\alpha;\mbox{\boldmath$\tau$}_{j}\ell m}^{*}\left(\mathbf{r}\right)\psi_{n\mathbf{k}\alpha}\left(\mathbf{r}\right)d^{3}r.\label{eq:coefficient_def}\end{equation}
Here, function $\theta\left(x\right)$ is defined that it is unity
when $x>0$ and zero other than the case. $Y_{\ell m}\left(\mathbf{e}\right)$
is spherical harmonics $Y_{\ell m}\left(\theta,\phi\right)$ with
$e_{x}=\sin\theta\cos\phi$, $e_{y}=\sin\theta\sin\phi$, and $e_{z}=\cos\theta$.
Wave function $\psi_{n\mathbf{k}\alpha}\left(\mathbf{r}\right)$ is
for the state specified by band index $n$, wave vector $\mathbf{k}$,
and spin $\alpha$ with energy $\epsilon_{n\mathbf{k}\alpha}$; vector
$\mathbf{r}$ represents the relative position from $\mbox{\boldmath$\tau$}_{j}$.
$\epsilon_{1s\alpha}$ represent the $1s$ core level with spin $\alpha$,
and $\Gamma$ is the core hole lifetime broadening. Wave function
$\psi_{n\mathbf{k}\alpha;\mbox{\boldmath$\tau$}_{j}\ell m}\left(\mathbf{r}\right)$
is the projection of $\psi_{n\mathbf{k}\alpha}\left(\mathbf{r}\right)$
on the state specified by azimuthal quantum number $\ell$ and magnetic
quantum number $m$ inside the MT sphere at $\mbox{\boldmath$\tau$}_{j}$
(it is normalized inside the MT sphere). We write the wave function
as \begin{equation}
\psi_{n\mathbf{k}\alpha;\mbox{\boldmath$\tau$}_{j}\ell m}\left(\mathbf{r}\right)=R_{n\mathbf{k}\alpha;\mbox{\boldmath$\tau$}_{j}\ell m\alpha}\left(r\right)Y_{\ell m}\left(\hat{\mathbf{r}}\right),\label{eq:local_orbital_state}\end{equation}
where $R_{n\mathbf{k}\alpha;\mbox{\boldmath$\tau$}_{j}\ell m\alpha}\left(r\right)$
is the normalized radial wave function.

The dipole transition matrix element $T_{n\mathbf{k}\alpha;\mbox{\boldmath$\tau$}_{j}1m\alpha}$
is given by

\begin{eqnarray}
 &  & T_{n\mathbf{k}\alpha;\mbox{\boldmath$\tau$}_{j}1m\alpha}\nonumber \\
 &  & =\int_{0}^{R_{\mathrm{MT}}}R_{n\mathbf{k}\alpha;\mbox{\boldmath$\tau$}_{j}1m\alpha}\left(r\right)\frac{r}{a_{0}}R_{\mbox{\boldmath$\tau$}_{j}1s\alpha}\left(r\right)\! r^{2}dr,\label{eq:dipole_mat}\end{eqnarray}
where $R_{\mbox{\boldmath$\tau$}_{j}1s\alpha}\left(r\right)$ is the
radial wave function of the $1s$ state with spin $\alpha$ at $\mathbf{\mbox{\boldmath$\tau$}}_{j}$.
The factor $A^{\prime}$ in Eq. (\ref{eq:formfac_reso}) is\begin{equation}
A^{\prime}=\frac{\alpha^{2}}{4}\left(\frac{\epsilon_{n\mathbf{k}\alpha}-\epsilon_{1s}}{E_{\mathrm{Ryd}}}\right)^{2}.\label{eq:prefac_reso}\end{equation}
Since $\hbar\omega\approx\epsilon_{n\mathbf{k}\alpha}-\epsilon_{1s}$,
this factor nearly equal to the factor $A$ in the nonresonant scattering
amplitude. Hereafter, we replace $A^{\prime}$ by $A$.

We rewrite the site scattering factor $f_{\mathrm{OO}}^{\mbox{\boldmath$\tau$}_{j}}$
as

\begin{eqnarray}
 &  & f_{\mathrm{OO}}^{\mbox{\boldmath$\tau$}_{j}}\left(\mathbf{q}^{\prime},\mathbf{e}^{\prime}\leftarrow\mathbf{q},\mathbf{e}\right)\nonumber \\
 &  & =A\frac{4\pi}{9}\sum_{m}\sum_{m^{\prime}}Y_{1m^{\prime}}\left(\mathbf{e}^{\prime}\right)Y_{1m}^{*}\left(\mathbf{e}\right)\nonumber \\
 &  & \times\int_{-\infty}^{\infty}\!\!\! d\epsilon\frac{\rho_{m^{\prime}m}^{\mathbf{\mbox{\boldmath$\tau$}}_{j}}\left(\epsilon\right)}{\left(\hbar\omega-\epsilon-\epsilon_{1s\alpha}+i\Gamma\right)/mc^{2}}\theta\left(\epsilon-\epsilon_{\mathrm{F}}\right),\label{eq:form_reso_2}\end{eqnarray}
where the dipole transition density matrix $\rho_{m^{\prime}m}^{\mathbf{\mbox{\boldmath$\tau$}}_{j}}\left(\epsilon\right)$
is defined by\begin{eqnarray}
 &  & \rho_{m^{\prime}m}^{\mathbf{\mbox{\boldmath$\tau$}}_{j}}\left(\epsilon\right)=\sum_{\alpha n\mathbf{k}}T_{n\mathbf{k}\alpha;\mbox{\boldmath$\tau$}_{j}1m^{\prime}\alpha}^{\dagger}T_{n\mathbf{k}\alpha;\mbox{\boldmath$\tau$}_{j}1m\alpha}\nonumber \\
 &  & \qquad\times F_{n\mathbf{k}\alpha}^{\mbox{\boldmath$\tau$}_{j}1m^{\prime}}F_{n\mathbf{k}\alpha}^{\mbox{\boldmath$\tau$}_{j}1m*}\delta\left(\epsilon-\epsilon_{n\mathbf{k}\alpha}\right).\label{eq:tran_den_mat}\end{eqnarray}
Incidentally, the absorption intensity $I_{\mathrm{abs}}$ in the
dipole process may be given by\begin{equation}
I_{\mathrm{abs}}\sim-\Im\sum_{jm}\int_{-\infty}^{\infty}\!\!\! d\epsilon\frac{\rho_{mm}^{\mathbf{\mbox{\boldmath$\tau$}}_{j}}\left(\epsilon\right)}{\left(\hbar\omega-\epsilon-\epsilon_{1s\alpha}+i\Gamma\right)/mc^{2}},\label{eq:abs_int}\end{equation}
where $\Im A$ stands for the imaginary part of $A$.

The matrix $\rho_{m^{\prime}m}^{\mathbf{\mbox{\boldmath$\tau$}}_{j}}\left(\epsilon\right)$
bears the symmetry compatible with the LSDW or TSDW states. Let $\rho^{\mathbf{\mbox{\boldmath$\tau$}}_{j}}\left(\epsilon\right)$
be\begin{equation}
\rho^{\mathbf{\mbox{\boldmath$\tau$}}_{j}}\left(\epsilon\right)=\left(\begin{array}{rrr}
a & \alpha & \gamma\\
\alpha^{*} & b & \beta\\
\gamma^{*} & \beta^{*} & c\end{array}\right).\label{eq:tra_mat_den00}\end{equation}
Then $\rho^{\mathbf{\mbox{\boldmath$\tau$}_{j}}+10a\hat{\mathbf{z}}(\hat{\mathbf{y}})}\left(\epsilon\right)$
has to be\begin{equation}
\rho^{\mathbf{\mbox{\boldmath$\tau$}}_{j}+10a\hat{\mathbf{z}}(\hat{\mathbf{y}})}\left(\epsilon\right)=\left(\begin{array}{rrr}
c & -\beta^{*} & \gamma^{*}\\
-\beta & b & -\alpha^{*}\\
\gamma & -\alpha & a\end{array}\right),\label{eq:tra_mat_den10}\end{equation}
in the LSDW(TSDW) state with $Q_{\mathrm{LSDW(TSDW)}}=\frac{2\pi}{a}\frac{19}{20}$.
Note that the off-diagonal elements $\alpha$, $\beta$ and $\gamma$
are zero in the LSDW state, because of the four fold rotational symmetry
along the $z$ direction. In the TSDW state, this symmetry is not
hold in a strict sense. However, one may expect that it is approximately
hold, because the wave length of the SDW is quite long $\lambda_{{\rm SDW}}=20a$,
leading to very small off-diagonal elements in the transition density
matrix.

The scattering amplitude for the resonant process may be rewritten
as \begin{eqnarray}
 &  & f_{\mathrm{R}}\left(\mathbf{q}^{\prime},\mathbf{e}^{\prime}\leftarrow\mathbf{q},\mathbf{e}\right)\nonumber \\
 &  & =\frac{4\pi}{9}ANG\left(\mathbf{Q}\right)\sum_{m^{\prime}m}Y_{1m^{\prime}}\left(\mathbf{e}^{\prime}\right)Y_{1m}^{*}\left(\mathbf{e}\right)\nonumber \\
 &  & \times\int_{-\infty}^{\infty}\!\!\! d\epsilon\frac{\tilde{\rho}_{m^{\prime}m}\left(\epsilon,\mathbf{Q}\right)}{\left(\hbar\omega-\epsilon-\epsilon_{1s\alpha}+i\Gamma\right)/mc^{2}}\theta\left(\epsilon-\epsilon_{\mathrm{F}}\right),\label{eq:amp_reso2}\end{eqnarray}
where\begin{eqnarray}
\tilde{\rho}\left(\epsilon,\mathbf{Q}\right) & = & \sum_{j}e^{-i\mathbf{Q}\cdot\mathbf{\mbox{\boldmath$\tau$}_{j}}}\rho^{\mathbf{\mbox{\boldmath$\tau$}_{j}}}\left(\epsilon\right).\label{eq:tra_mat_den_cell0}\end{eqnarray}
Then $\tilde{\rho}\left(\epsilon,\mathbf{Q}\right)$ has the form\begin{equation}
\tilde{\rho}\left(\epsilon,\mathbf{Q}\right)=\left(\begin{array}{rrr}
\xi & \eta & \zeta\\
\eta^{*} & 0 & \eta^{*}\\
\zeta^{*} & \eta & -\xi\end{array}\right).\label{eq:tra_mat_den_cell}\end{equation}
From the above-mentioned symmetry argument on $\rho^{\mathbf{\mbox{\boldmath$\tau$}}_{j}}\left(\epsilon\right)$,
the off-diagonal elements $\eta$,$\zeta$ are zero in the LSDW state.
The scattering amplitude for the $\sigma\pi^{\prime}$ channel in
the LSDW state becomes\begin{eqnarray}
 &  & f_{\mathrm{R}}\left(\mathbf{q}^{\prime},\mathbf{e}^{\pi\prime}\leftarrow\mathbf{q},\mathbf{e}^{\sigma}\right)=-\frac{1}{3}ANG\left(\mathbf{Q}\right)i\sin\theta\nonumber \\
 &  & \times\int_{-\infty}^{\infty}\!\!\! d\epsilon\frac{\xi\left(\epsilon,\mathbf{Q}\right)}{\left(\hbar\omega-\epsilon-\epsilon_{1s\alpha}+i\Gamma\right)/mc^{2}}.\label{eq:amp_reso_lsdw}\end{eqnarray}
It is easily seen that the amplitude for the $\sigma\sigma^{\prime}$
channel is zero. Also, in the TSDW state, the off-diagonal elements
$\eta$, $\zeta$ are expected to be very small. Neglecting them,
we obtain the scattering amplitude for the $\sigma\pi^{\prime}$ channel
in the TSDW state as\begin{eqnarray}
 &  & f_{\mathrm{R}}\left(\mathbf{q}^{\prime},\mathbf{e}^{\pi\prime}\leftarrow\mathbf{q},\mathbf{e}^{\sigma}\right)=\frac{1}{6}ANG\left(\mathbf{Q}\right)i\nonumber \\
 &  & \times\int_{-\infty}^{\infty}\!\!\! d\epsilon\Biggl[\frac{\left(2\xi-\zeta+\zeta^{*}\right)\cos\theta\sin\psi}{\left(\hbar\omega-\epsilon-\epsilon_{1s\alpha}+i\Gamma\right)/mc^{2}}\nonumber \\
 &  & \quad-\frac{\sqrt{2}\left(\eta+\eta^{*}\right)\cos\theta\cos\psi}{\left(\hbar\omega-\epsilon-\epsilon_{1s\alpha}+i\Gamma\right)/mc^{2}}\Biggr]\theta\left(\epsilon-\epsilon_{\mathrm{F}}\right)\nonumber \\
 &  & \approx\frac{1}{3}ANG\left(\mathbf{Q}\right)i\cos\theta\sin\psi\nonumber \\
 &  & \times\int_{-\infty}^{\infty}\!\!\! d\epsilon\frac{\xi\left(\epsilon,\mathbf{Q}\right)}{\left(\hbar\omega-\epsilon-\epsilon_{1s\alpha}+i\Gamma\right)/mc^{2}}.\label{eq:amp_reso_tsdw}\end{eqnarray}
 The amplitude in the $\sigma\sigma^{\prime}$ is zero if the off-diagonal
elements $\eta$, $\zeta$ are zero.

Using the non-resonant and resonant scattering amplitude thus obtained,
the scattering cross section can be written as,\begin{equation}
\frac{d\sigma}{d\Omega}=r_{e}^{2}\left|f_{\mathrm{NR}}+f_{\mathrm{R}}\right|^{2},\label{eq:cross_section}\end{equation}
where $r_{e}=\frac{e^{2}}{mc^{2}}$ represents the classical electron
radius.

\section{Results and Discussion}

\subsection*{Absorption spectra}

\begin{figure}[h]
\begin{center}\includegraphics[%
  scale=0.7]{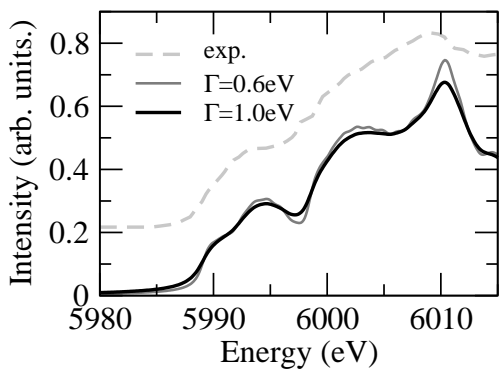}

\caption{Comparison of the absorption spectra. Solid and broken lines represent
the calculated spectra and experiment. $1s$ core level is assumed
to be $\epsilon_{1s}=\epsilon_{\mathrm{F}}-5988.4\,\mathrm{eV}$.
The lifetime broadening of the intermediate states is assumed to be
$\Gamma=0.6\,\mathrm{eV}$ and $1.0\,\mathrm{eV}$.\label{cap:Comparison-of-the}}

\end{center}
\end{figure}

At present, it is difficult to estimate accurately the $1s$ core
level $\epsilon_{1s}$ and the core-hole lifetime broadening $\Gamma$
from the calculation. We adjust these values such that the calculated
$K$-edge absorption intensity reproduces the photon energy dependence
observed in the experiment.\cite{Mannix2001} We calculate the spectra
using Eq. (\ref{eq:abs_int}), assuming that $\epsilon_{1s}=\epsilon_{\mathrm{F}}-5988.4\,\mathrm{eV}$
and $\Gamma=0.6\sim1\,\mathrm{eV}$. Figure \ref{cap:Comparison-of-the}
shows the calculated absorption spectra in comparison with the experiment.
The calculated spectral shape is roughly proportional to the PDOS
for the unoccupied $4p$ state. The calculated curves agree well with
the experiment except for a shape around $\hbar\omega=6010\,\mathrm{eV}$.
The peak around $6010\,\mathrm{eV}$ comes from the peak around $21\,\mathrm{eV}$
in the $p$ PDOS. Note that no noticeable difference is found between
the LSDW and TSDW states. This good agreement between the calculation
and the experiment may justify the neglect of the core-hole potential
in the calculation of resonant scattering spectra.

\subsection*{magnetic scattering spectra}

\begin{figure}[h]
\begin{center}\includegraphics[%
  clip,
  scale=0.7]{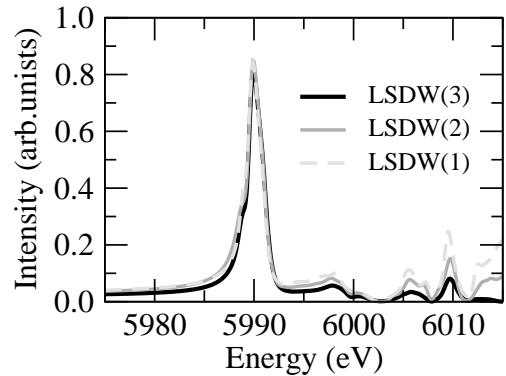}

\caption{Convergence of the scattering intensity at $\mathbf{Q}=\mathbf{Q}_{\mathrm{LSDW}}$
for LSDW state against the number of $k$-points. The number of $k$-points
are $12\times12\times2$, $16\times16\times2$ and $30\times30\times2$
for LSDW(1), LSDW(2) and LSDW(3), respectively. The $1s$ core level
and the lifetime broadening assumed to be $\epsilon_{1s}=\epsilon_{\mathrm{F}}-5988.4\,\mathrm{eV}$
and $\Gamma=0.6\,\mathrm{eV}$, respectively. \label{cap:rxs_int_kdep}}

\end{center}
\end{figure}

Putting $\Gamma=0.6\,\mathrm{eV}$ and using the same values of $\epsilon_{1s}$
as in the absorption spectra, we calculate the magnetic scattering
spectra. We first check the convergence of the spectral curve with
varying number of $k$-points $n_{k}$ in the FBZ. Figure \ref{cap:rxs_int_kdep}
shows the spectral curves calculated with three values of $n_{k}$
at $\mathbf{Q}=\mathbf{Q}_{\mathrm{LSDW}}$ in the LSDW state. The
spectra calculated at $\hbar\omega>6000\,\mathrm{eV}$ show strong
dependence on $n_{k}$, no sign of convergence even for $n_{k}=30\times30\times2$.
On the other hand, the curve around $\hbar\omega=5990\ \mathrm{eV}$
depends little on $n_{k}$, suggesting a convergence even for $n_{k}=12\times12\times2$.

\begin{figure}[h]
\begin{center}\includegraphics[%
  clip,
  scale=0.7]{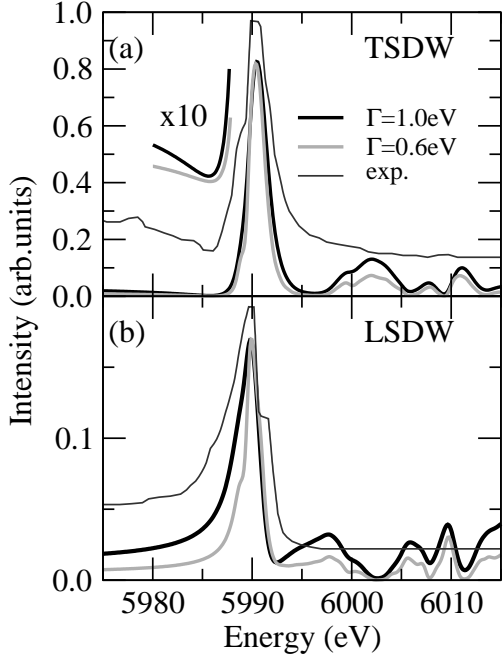}

\caption{(a) Calculated magnetic scattering intensities for $\sigma\pi^{\prime}$
channel on the SDW Bragg spots $\mathbf{Q}_{\mathrm{TSDW}}$ in the
TSDW state. The azimuthal angle is assumed to be $\psi=90\,\mathrm{deg.}$
(b) Those on the spot $\mathbf{Q}_{\mathrm{LSDW}}$ in the LSDW state.
The number of $k$ points in FBZ is $n_{k}=16\times16\times2$. The
azimuthal angle is assumed to be $\psi=\frac{\pi}{2}$ for the TSDW
state. $1s$ core level is assumed to be $\epsilon_{1s}=\epsilon_{\mathrm{F}}-5988.4\,\mathrm{eV}$.
Both results with $\Gamma=0.6\,\mathrm{eV}$ and $\Gamma=1.0\,\mathrm{eV}$
are shown for comparison. The spectra with $\Gamma=1.0\,\mathrm{eV}$
is normalized so that the peak hight at $5990\,\mathrm{eV}$ is equal
with that of the spectra with $\Gamma=0.6\,\mathrm{eV}$. The experiment
curve is traced from Ref.\cite{Mannix2001}.\label{cap:rxs_int_lsdw}}

\end{center}
\end{figure}

Figure \ref{cap:rxs_int_lsdw} shows the calculated spectral curves
in the LSDW and TSDW states, in comparison with the experiment\cite{Mannix2001}.
The absorption correction is not made in the experimental curves.
The calculation reproduces successfully not only the Fano-type dip
around $\hbar\omega=5986\,\mathrm{eV}$ in the TSDW state but also
the resonant behavior in good agreement with the experiment. 

\begin{figure}[h]
\begin{center}\includegraphics[%
  clip,
  scale=0.7]{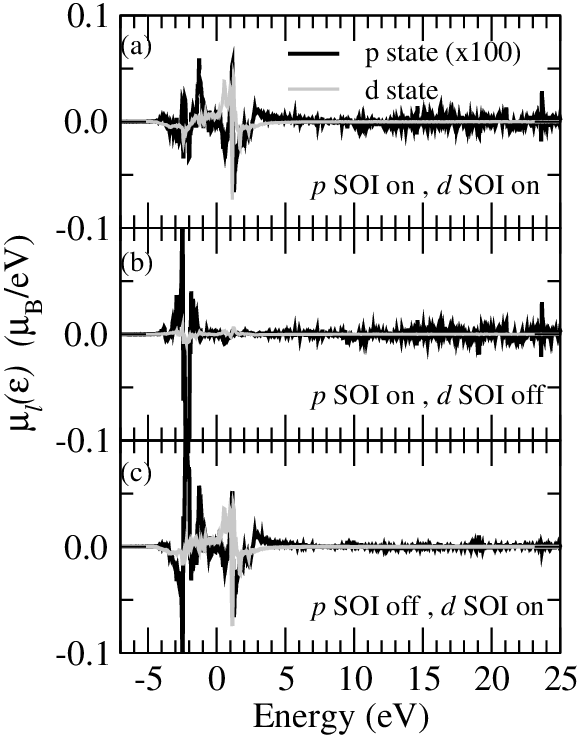}

\caption{Orbital moment densities $\mu_{1}^{\mbox{\boldmath$\tau$}_{j}}\left(\epsilon\right)$
and $\mu_{2}^{\mbox{\boldmath$\tau$}_{j}}\left(\epsilon\right)$ on
the site at $\mbox{\boldmath$\tau$}_{0}$. (a) Both of the $p$ SOI
and $d$ SOI are turned on. (b) The $d$ SOI is turned off. (c) The
$p$ SOI is turned off. Solid line and gray line represent the orbital
density for the $p$ state $\mu_{1}^{\mbox{\boldmath$\tau$}_{j}}\left(\epsilon\right)$
and $d$ symmetric states $\mu_{2}^{\mbox{\boldmath$\tau$}_{j}}\left(\epsilon\right)$
, respectively. The orbital density for the $p$ states is multiplied
by $100$. The number of $k$-points in FBZ is $30\times30\times2$.
\label{cap:Orbital-moment-density.}}

\end{center}
\end{figure}

The resonant intensity arises from the OP in the $4p$ bands. In order
to clarify how the OP is induced, we investigate the change of the
electronic states by turning on and off the SOI on the $p$ and $d$
states selectively. Figure \ref{cap:Orbital-moment-density.} shows
the density of OP $\mu_{\ell}^{\mbox{\boldmath$\tau$}_{0}}\left(\epsilon\right)$
as a function of energy in the LSDW state. This quantity is defined
as\begin{equation}
\mu_{\ell}^{\mbox{\boldmath$\tau$}_{0}}\left(\epsilon\right)=\sum_{m=-\ell}^{\ell}m\rho_{\ell m}^{\mbox{\boldmath$\tau$}_{0}}\left(\epsilon\right),\label{eq:orbital_density}\end{equation}
where $\rho_{\ell m}^{\mbox{\boldmath$\tau$}_{0}}\left(\epsilon\right)$
is the partial DOS projected on the state specified by azimuthal quantum
number $\ell$ and magnetic quantum number $m$ inside the MT sphere
at site $\mbox{\boldmath$\tau$}_{0}$. (The site $\mbox{\boldmath$\tau$}_{0}$
has the largest magnetic moment.) When the SOI is working on both
the $p$ and $d$ states (panel a), $\mu_{2}^{\mbox{\boldmath$\tau$}_{0}}\left(\epsilon\right)$
for the $d$ states is quite large for $0\lesssim\epsilon\lesssim2\,\mathrm{eV}$
but nearly zero for $\epsilon\gtrsim3\,\mathrm{eV}$. $\mu_{1}^{\mbox{\boldmath$\tau$}_{0}}\left(\epsilon\right)$
for the $p$ states is smaller in two order of magnitude than $\mu_{2}^{\mbox{\boldmath$\tau$}_{0}}\left(\epsilon\right)$.
When the $d$ SOI is turned off but the $p$ SOI is still turned on
(panel b), $\mu_{2}^{\mbox{\boldmath$\tau$}_{0}}\left(\epsilon\right)$
becomes very small in the whole energy range. Corresponding to this
change, $\mu_{1}^{\mbox{\boldmath$\tau$}_{0}}\left(\epsilon\right)$
also becomes very small for $0\lesssim\epsilon\lesssim5\,\mathrm{eV}$,
although it does not change for $\epsilon\gtrsim5\,\mathrm{eV}$ from
the values in the case when both of the $p$ and $d$ SOI's are turned
on. When only the $p$ SOI is turned off but the $d$ SOI is turned
on (panel c), $\mu_{2}^{\mbox{\boldmath$\tau$}_{0}}\left(\epsilon\right)$
is essentially unchanged in the whole energy range from the value
in the case when both of the $p$ and $d$ SOI's are turned on. $\mu_{1}^{\mbox{\boldmath$\tau$}_{0}}\left(\epsilon\right)$
changes little for $\epsilon\lesssim5\,\mathrm{eV}$, but it is considerably
reduced for $\epsilon\gtrsim10\,\mathrm{eV}$.

These results clearly indicate that the $p$ OP for $0\lesssim\epsilon\lesssim5\,\mathrm{eV}$
is mainly induced by the $d$ OP through the $p$-$d$ hybridization.
Since the $d$ DOS is concentrated in $\epsilon\lesssim2\,\mathrm{eV}$,
the hybridization with the $p$ states at neighboring site may be
effective only for $\epsilon\lesssim2\,\mathrm{eV}$. The effect of
the $p$ SOI on the $p$ OP seems to be reduced by this hybridization.
Note that the $p$ and $d$ states do not hybridize each other at
the same site because the SDW wavelength is so long that each site
almost keeps the inversion symmetry, particularly at $\mbox{\boldmath$\tau$}_{0}$.
The $p$ OP for $\epsilon\gtrsim10\,\mathrm{eV}$ is mainly induced
by the $p$ SOI. 

\begin{figure}[h]
\begin{center}

\includegraphics[%
  clip,
  scale=0.7]{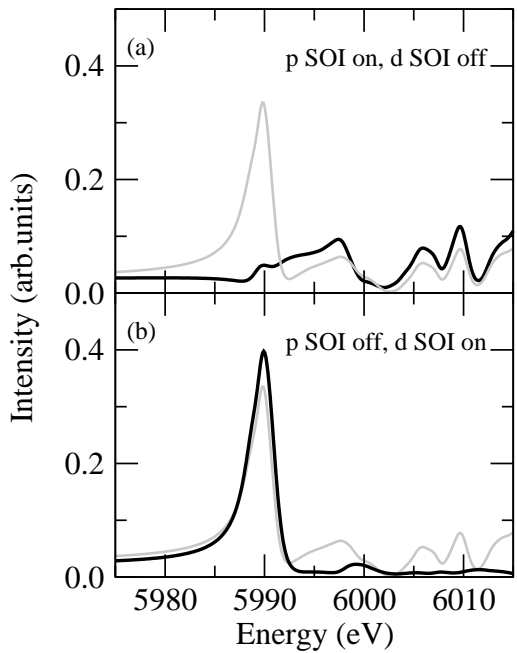}

\caption{Scattering intensity on the SDW spot $\mathbf{Q}_{\mathrm{LSDW}}$
in LSDW state. (a) The SOI in $p$ state is turned on and the SOI
in $d$ state is turned off. (b) The SOI in $p$ state is turned off
and the SOI in $d$ state in turned on. The $1s$ core level is assumed
as $\epsilon_{1s}=\epsilon_{\mathrm{F}}-5988.4\,\mathrm{eV}$. The
gray line is shown for comparison, which is calculated with both of
the $p$ SOI and $d$ SOI being turned on.\label{cap:Scat_on_off}}

\end{center}
\end{figure}

Since the resonant intensity arises from the $p$ OP, the above finding
that the $p$ OP for $\epsilon\lesssim5\,\mathrm{eV}$ is induced
by the $d$ OP through the $p$-$d$ hybridization may be confirmed
by calculating directly the spectra with turning on and off the $p$
SOI and $d$ SOI. Figure \ref{cap:Scat_on_off} shows the calculated
results. When only the $d$ SOI is turned off, the resonant main peak
at $\hbar\omega=5990\,\mathrm{eV}$ disappears, while the resonant
structures for higher energies change little. On the other hand, when
only the $p$ SOI is turned off, the main peak dose not change, while
the resonant structure for higher energies almost disappears. This
SOI dependence of the spectra corresponds well to the SOI dependence
of the $p$ OP. Consequently, it is conclude that the resonant main
peak at $\hbar\omega=5990\,\mathrm{eV}$ strongly correlates with
the $3d$ states.

\begin{figure}[h]
\begin{center}\includegraphics[%
  clip,
  scale=0.7]{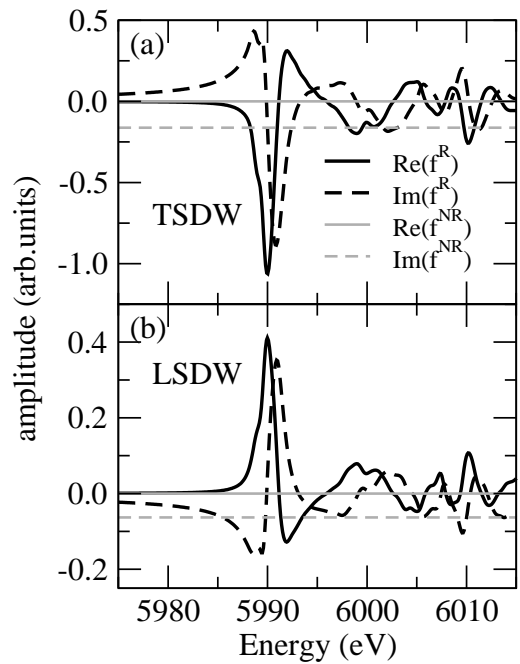}

\caption{(a) Scattering amplitudes on the spot $\mathbf{Q}_{\mathrm{TSDW}}$
for the TSDW state. (b) Those on the spot $\mathbf{Q}_{\mathrm{LSDW}}$
for the LSDW state. Solid and broken lines represent the real and
imaginary part of the amplitude due to the resonant scattering, respectively.
Thin broken line represents the imaginary part of the amplitude due
to the nonresonant scattering. The real part of the nonresonant contributions
is zero. $\epsilon_{1s}=\epsilon_{\mathrm{F}}-5988.4\,\mathrm{eV}$
and $\Gamma=1.0\,\mathrm{eV}$.\label{cap:Scat_amp}}

\end{center}
\end{figure}

Finally we discuss the origin of the Fano-type dip. Since the Fano
effect arises from the interference between the non-resonant and resonant
terms, we look into $f_{\mathrm{NR}}$ and $f_{\mathrm{R}}$, which
are shown in Fig. \ref{cap:Scat_amp}. $f_{\mathrm{NR}}$ is purely
imaginary. Note that $f_{\mathrm{R}}$ in the LSDW state is nearly
proportional to that in the TSDW state with reverse sign. The reverse
sign arises from the difference of the scattering geometry in the
LSDW and TSDW states. As a result, in the TSDW state, the imaginary
part of $f_{\mathrm{R}}$ and $f_{\mathrm{NR}}$ can be canceled at
$\hbar\omega\approx5987\,\mathrm{eV}$, with a small real part of
$f_{\mathrm{R}}$. This is the reason why the Fano dip is seen at
$\hbar\omega\approx5987\,\mathrm{eV}$. On the other hand, one can
easily see that such a cancellation is hard in the LSDW state.

\section{Concluding Remarks}

We have performed an \emph{ab initio} band structure calculation based
on the LDA with a large unit cell. Assuming that the potential has
a modulation with $Q_{\mathrm{SDW}}=\frac{2\pi}{a}\frac{19}{20}$,
we reproduced well the SDW state previously reported.\cite{Hirai1998,Hafner2002}
Since the orbital momentum is essential to the resonant magnetic scattering,
we have taken account of the SOI within the $\ell_{z}s_{z}$ approximation.
The orbital magnetic moment is mainly induced on the $3d$ states
in proportion to the spin magnetic moment at each Cr site.

We have calculated the absorption spectra at the Cr $K$-edge. The
dipole matrix elements are evaluated by using the wave functions obtained
from the band structure calculation. Having neglected the core-hole
potential, we have obtained the absorption spectra in good agreement
with the experiment.\cite{Mannix2001} This may justify neglecting
the core-hole potential in the calculation of the resonant magnetic
scattering spectra.

We have calculated the resonant x-ray magnetic scattering spectra
on the SDW magnetic Bragg spot in the TSDW and LSDW states. A resonant
enhancement has been reproduced near the $K$-edge in good agreement
with the experiment.\cite{Mannix2001} In the TSDW state, the off-diagonal
matrix elements of the scattering amplitude have been neglected, since
they are expected to be small from a symmetry argument. Experiments
on the azimuthal angle dependence of the resonant intensity may help
us check the accuracy of the argument.

The mechanism of the resonant magnetic scattering has been clarified.
We found that the OP on the $3d$ states fluctuates considerably as
a function of energy, although the the local orbital magnetic moment
in the $d$ states is quite small. This $d$ OP induces the $p$ OP
around the Fermi level through the $p$-$d$ hybridization, and thereby
gives rise to the resonant scattering intensity. The resonant main
peak at $\hbar\omega=5990\,\mathrm{eV}$ strongly correlates with
the $3d$ states. This finding is consistent with the analysis\cite{Usuda2004}
of the resonant magnetic scattering of UGa$_{3}$\cite{Mannix2001b},
where a large signal is found at the Ga $K$-edge. This is interpreted
as a reflection of the large OP on the uranium $5f$ states. 

We have reproduced a Fano-type dip at $\hbar\omega\approx5987\,\mathrm{eV}$
in the TSDW state.\cite{Mannix2001} The Fano-type dip is seen only
in the TSDW state. This difference arises from the sign of the resonant
scattering amplitude (the energy dependence of the amplitude is nearly
the same in the two states except for the sign), owing to the different
scattering geometry. 

Finally we comment on the CDW state. The assumption of an ideal bcc
lattice gives rise to only a small CDW amplitude. The LSW may be dispensable
for obtaining a sufficient magnitude of the CDW amplitude. A detailed
study on the CDW state is left in future. 

\begin{acknowledgments}
This research was partially supported by the Ministry of Education,
Science, Sports and Culture, Japan, Grant-in-Aid for Young Scientists
(B), 15740196, 2003 and Grant-in-Aid for Scientific Research (C)(2).
\end{acknowledgments}
\bibliographystyle{apsrev}
\bibliography{paper_us2}

\end{document}